\documentclass[onecolumn,pre,amsmath,amssymb]{revtex4}

\usepackage{amsmath}
\usepackage{amssymb,relsize}
\usepackage[dvips]{graphicx}

\usepackage{epsf}

\newcommand{\beq}{\begin{equation}}
\newcommand{\eeq}{\end{equation}}
\newcommand{\beqa}{\begin{eqnarray}}
\newcommand{\eeqa}{\end{eqnarray}}
\newcommand{\beqan}{\begin{eqnarray*}}
\newcommand{\eeqan}{\end{eqnarray*}}
\newcommand{\ben}{\begin{enumerate}}
\newcommand{\een}{\end{enumerate}}
\newcommand{\bit}{\begin{itemize}}
\newcommand{\eit}{\end{itemize}}

\newcommand{\refeq}[1]{(\ref{#1})}
\newcommand{\code}[1]{{\smaller\texttt{#1}}}
\newcommand{\ep}{\thinspace . }

\newcommand{\mathbfx}[1]{{\mbox{\boldmath $#1$}}}

\newcommand{\modten}[1]{\mathbfx{#1}}
\newcommand{\modvec}[1]{\mathbfx{#1}}
\newcommand{\phivec}{\mathbfx{\phi}}

\renewcommand{\u}{\mathbfx{u}}
\renewcommand{\b}{\mathbfx{b}}

\newcommand{\K}{\mathbfx{K}}

\renewcommand{\L}{\mathbfx{L}}

\newcommand{\D}{\mathbfx{D}}
\newcommand{\M}{\mathbfx{M}}

\newcommand{\bft}{\mathbfx{t}}
\newcommand{\B}{\mathbfx{B}}
\newcommand{\Q}{\mathbfx{Q}}
\renewcommand{\P}{\mathbfx{P}}

\newcommand{\p}{\mathbfx{p}}

\newcommand{\f}{\mathbfx{f}}

\newcommand{\tde}{\modten{\varepsilon}}
\newcommand{\tds}{\modten{\sigma}}

\begin{document}

\title{A finite element method for modelling electromechanical wave propagation in anisotropic
piezoelectric media}
\newcommand{\gatefont}[1]{\textbf{#1}}

\author{S. Rahman$^1$}
\email{S.Rahman.00@cantab.net}
% \altaffiliation[Also at ]{Physics Department, Cambridge University.}%Lines break automatically or can be forced with \\
\author{H. P. Langtangen$^2$}%
\author{C. H. W. Barnes$^1$}%
\affiliation{$^1$Cavendish Laboratory, Cambridge
University, J J Thomson Avenue, Cambridge, CB3 OHE, United Kingdom}%
\affiliation{$^2$Simula Research Laboratory, Martin Linges v 17,
Fornebu P.O.Box 134, 1325 Lysaker, Norway}%

%\tableofcontents

\begin{abstract}

We describe and evaluate a numerical solution strategy for
simulating surface acoustic waves (SAWs) through semiconductor
devices with complex geometries. This multi-physics problem is of
particular relevance to the design of SAW-based quantum electronic
devices. The mathematical model consists of two coupled partial
differential equations for the elastic wave propagation and the
electric field, respectively, in anisotropic piezoelectric media.
These equations are discretized by the finite element method in
space and by a finite difference method in time. The latter method
yields a convenient numerical decoupling of the governing
equations.  We describe how a computer implementation can utilize
the decoupling and, via object-oriented programming techniques
reuse independent codes for the Poisson equation and the linear
time-dependent elasticity equation.  First we apply the simulator
to a simplified model problem for verifying the implementation,
and thereafter we show that the methodology is capable of
simulating a real-world case from nanotechnology, involving SAWs
in a geometrically non-trivial device made of Gallium Arsenide.
\end{abstract}

\pacs{02.70.Dc, 85.35.Gv, 73.23.-b, 77.65.Dq, 02.60.–x}

\maketitle \emph {}

%Keywords : Finite Element Method, Piezoelectric Surface Acoustic
%Waves, Gallium Arsenide.

\section{Introduction}

In the process of designing quantum electronic devices based on
surface acoustic waves (SAWs) traversing piezoelectric media, it
is necessary to determine the effect, on these waves of obstacles
such as electrical gates on the surface, and also the effect of
the SAW on the low-dimensional quantum mechanical systems such as
quantum wires and quantum dots. In general, the gates have a
non-trivial geometry, which necessitate numerical simulation
tools.  The finite element method is well suited to handle complex
geometries and is widely used to model piezoelectric devices
\cite{PiezoSimLerch1,Benjeddou:00,Ha,Saravanos:99,Gopinathan:00,Makerle:01}.
The effect on the low-dimensional quantum mechanical systems would
be analyzed through coupling the SAW simulator to both stationary
and time-dependent Schr\"{o}dinger equations \cite{Mohan}. This
would require the development of a fast SAW simulator but also
flexible and portable code.

SAWs are modes of propagation of energy along the free surface of
a material such that there is no decay along the direction of
propagation but there is exponential decay into the bulk. SAWs
have been used in electrical sensor technologies for many decades
and have also been a useful tool in probing quantum electronic
structures, for example, quantum Hall liquids \cite{Wixforth:86}.
Recently, much experimental work has been done in the field of
acoustic charge transport whereby a SAW across a GaAs/AlGaAs
heterostructure is used to capture a single electron and then
transport it along a one-dimensional quantum wire \cite{Shilton1}.
This would be useful in developing an accurate current standard,
but more challenging proposals to use this in the burgeoning field
of quantum information processing have been proposed
\cite{BarnesQC1,BarnesQC2,Bertoni:00,Bordone:04,AcousticSinglePhotoSrc}.
SAWs have also been utilized for both static  quantum dot
\cite{Ebbecke:03} and photo-luminescence experiments
\cite{Alsini:03,Sogawa:01}. The time and resources required to
build such devices are immense, and therefore the mathematical
modelling of these devices before the physical construction is
advantageous. This approach requires the solution of the continuum
electromechanical equations of motion in a piezoelectric medium.
The method of partial waves \cite{Matthews} can be used to obtain
simple analytical expressions for the waves in the bulk material,
but the the solutions say nothing about the effect of gates on the
surface. Attempts to solve the governing equations analytically
for devices which do have surface gates
\cite{J.J.Campbell.Jones,AizinGumbs1,AizinGumbs2} involve
simplifications, and the accuracy of these approximations remains
uncertain.

There is a vast amount of literature on the three-dimensional
finite element analysis of piezoelectric devices
\cite{PiezoSimLerch1,Benjeddou:00,Ha,Saravanos:99,Gopinathan:00,Makerle:01},
and these methods are exploited in the field of ultrasonics to
design control systems involving piezoelectric actuators and
sensors \cite{Preumont}. Commercial software such as ANSYS and
ABAQUS can be used to simulate electromechanical phenomena but
lack the flexibility to couple to quantum mechanical calculations
such as iterative Poisson-Schr\"{o}dinger \cite{Mohan}. Therefore
it is often necessary to develop ones own computer code for
modelling such devices.

In this paper, we formulate and evaluate a finite element based
solution method for the equations governing SAWs in piezoelectric
media, and we formulate it in a flexible and portable manner to
allow the ability to interface with other code. The implementation
is performed in an object-oriented style in order to incorporate
existing solvers and to enhance portability of the code. Such a
tool can be valuable in the design of micro- and nano-scale
devices. We describe a set of boundary conditions that are capable
of efficiently exciting SAWs and demonstrate the propagation of
SAWs through a GaAs-based device. Although our applications are
specific to GaAs, the formulation is general enough to allow
simulations, with the same code (or with small modifications), of
any crystal structure provided the elastic and piezoelectric
material parameters are known.

This paper is organized as follows. In Section
\ref{sec:constitutiveRels} we give an overview of the mathematical
model underlying piezoelectricity and precisely state the coupled
partial differential equations we aim to solve. Sections
\ref{sec:FEformulation} and \ref{sec:optimisation} concern finite
element formulations of these equations and an overview of the
computational algorithm is described. In section
\ref{sec:Implementation} we describe the class based simulator
approach used in our code design which simplifies implementation
and increases reliability. Section \ref{sec:verification} presents
a verification of the implementation and an empirical estimation
of convergence properties of the numerical solution method. A
real-world application of the methodology, concerning a SAW
problem in nanotechnology is the subject of Section
\ref{sec:Application}, before we make some concluding remarks in
the final section.

\section{Modelling of piezoelectric materials}
\label{sec:constitutiveRels}

For piezoelectric materials, the electric displacement depends on
both the applied electric field and mechanical strain, and the
stresses depend on both the applied mechanical strain and applied
electric field. The electric displacement $D_i$ is given by

\begin{equation}\label{eqn:Di}
    D_i = \epsilon_{ij}^S E_j + e_{ijk} \varepsilon_{jk},
\end{equation}

\noindent where $E_j$ denotes the electric field, $\epsilon_{ij}$
is the permittivity tensor, $e_{ijk}$ is the piezoelectric
coupling constant, $\varepsilon_{ij}$ is the strain tensor. The
stress tensor $\sigma_{ij}$ is given by

\begin{equation}\label{eqn:Tij1}
    \sigma_{ij} = -e_{kij} E_k + c^E_{ijkl} \varepsilon_{kl},
\end{equation}

\noindent where  $c_{ijkl}$ represents the 4th rank tensor of
elastic parameters. Summation over repeated indices is implied in
this section,  and the superscripts $S$ and $E$ denote that the
quantities were measured under constant strain and constant
electric field, respectively.

In the absence of free charges within the material, Gauss's law
requires the divergence of the electric displacement to vanish:
\begin{equation}\label{eqn:DivD}
 \nabla \cdot D = \varrho_{\rm free} = 0 \ep
\end{equation}
Combining Eqs.~(\ref{eqn:Di}) and (\ref{eqn:Tij1}), and the
relation \beq E_i=-\frac{\partial \phi}{\partial x_i}
\label{eqn:Ei} \eeq between the electric potential $\phi$ and the
electric field $E_i$, together with the strain-displacement
relation for small strains,

\begin{equation}\label{eqn:Sij}
    \varepsilon_{ij} = \frac{1}{2} (\frac{\partial u_j}{\partial x_i} + \frac{\partial u_i}{\partial x_j}),
\end{equation}

\noindent where $u_i$ is the mechanical displacement field, we can
derive a scalar partial differential equation for $\phi$:

\begin{equation}\label{eqn:PiezoCoup2}
 \frac{\partial}{\partial x_i}\epsilon_{ik}^{S}\frac{\partial \phi}{ \partial
x_k} = \frac{\partial}{\partial x_i} e_{ikl} \frac{\partial
u_l}{\partial x_k}\ep
\end{equation}

This equation couples the potential $\phi$ and the displacements
$u_i$ in the medium, and can be used to compute the electric
field. Equation~\refeq{eqn:Ei} assumes the quasi-static
approximation, which requires that very little energy is carried
away by electromagnetic waves.

The displacement field in the medium is governed by
Newton's 2nd law of motion combined with the appropriate constitutive
law. The former equation reads
\begin{equation}\label{eqn:eqnMotion}
    \varrho \ddot{u}_i = \frac{\partial \sigma_{ij}}{\partial x_j} + \varrho b_i,
\end{equation}

\noindent where $b_i$ denotes body forces, and the double dot in
$\ddot u_i$ denotes a second-order partial derivative in time. As
Eq.~(\ref{eqn:eqnMotion}) stands, it contains no damping term,
which is irrelevant in our application setting. The constitutive
law stated in Eq.~(\ref{eqn:Tij1}) relates the stress tensor
$\sigma_{ij}$ to the strain tensor $\varepsilon_{ij}$, and
Eq.~\refeq{eqn:Sij} relates $\varepsilon_{ij}$ to the displacement
field $u_i$. Combining Eqs.~(\ref{eqn:Tij1}), (\ref{eqn:Sij}),
(\ref{eqn:eqnMotion}), we arrive at an equation for $u_i$ in terms
of $\phi$:

%and the symmetries of the elastic tensor,

%\begin{equation}\label{eqn:elasticSyms}
%c_{iklm} = c_{kilm} = c_{ikml} = c_{lmik},
%\end{equation}

\begin{equation}\label{eqn:dispField1}
\varrho \ddot{u}_i = \frac{\partial}{\partial
x_j}c_{ijkl}^E\frac{\partial u_l}{\partial x_k}   + \varrho b_i +  \frac{\partial}{\partial x_j}e_{kij}\frac{\partial \phi}{\partial x_k}\ep \\
\end{equation}
From Eq.~(\ref{eqn:dispField1}) we see how the mechanical motion
is affected the electric field. In most materials, this coupling
is weak, because $e_{ijk} \ll c_{ijkl}^E$, and the effect of the
piezoelectric potential on the mechanical deformation is
negligible. On the other hand, the effect of the mechanical
deformation on the potential is always significant according to
Eq.~\refeq{eqn:PiezoCoup2}. However, in cases where an external
electric field is applied through surface gates, the term in
$\phi$ may be sufficiently large to contribute significantly to
the mechanical motion, and these cases receive the focus of the
present paper.

%\label{sec:Mathematical Model} To summarize, the equations of motion
%for a piezoelectric medium are (\ref{eqn:PiezoCoup2}) and
%(\ref{eqn:dispField1}). The primary unknowns in these equations
%are $u_i$ and $\phi$, from which all other interesting quantities,
%such as stress, strain, and the electric field, can be derived
%through differentiation.

To illustrate the nature of these equations we explicitly write
out the coupling terms for the piezoelectric material GaAs. The
elastic parameters of a crystal are usually given in a coordinate
system with its $x$, $y$ and  $z$ axes parallel to the crystal X,
Y and Z axes. For consistency with acoustoelectric charge
transport experiments, we align the positive $x$ axis with the
crystal positive [011] axes, and the positive z axis with the
crystal positive [100] axis. To achieve this we perform a $45$
degree rotation of the crystal, around the $z$ axis.

In the transformed frame, the only non-zero components of the
piezoelectric tensor are $e_{15} = -e_{24} = e_{31} = -e_{32}$.
For generality, we illustrate our method using the different
values of the piezoelectric tensor. Using the assumptions stated
in the previous paragraph, and neglecting body forces, we can
write out equations Eqs.~(\ref{eqn:dispField1}) and
(\ref{eqn:PiezoCoup2}) as

\begin{widetext}
\begin{equation}\label{eqn:elastMotionx}
\varrho \ddot{u}_x = \frac{\partial}{\partial x_j} c_{xjkl}^E
\frac{\partial u_l}{\partial x_k} + \frac{\partial}{\partial x}
e_{31} \frac{\partial \phi}{\partial z} + \frac{\partial}{\partial
z} e_{15} \frac{\partial \phi}{\partial x},
\end{equation}

\begin{equation}\label{eqn:elastMotiony}
\varrho \ddot{u}_y = \frac{\partial}{\partial x_j} c_{yjkl}^E
\frac{\partial u_l}{\partial x_k} + \frac{\partial}{\partial y}
e_{32} \frac{\partial \phi}{ \partial z} +
\frac{\partial}{\partial z} e_{24} \frac{\partial \phi}{ \partial
y},
\end{equation}

\begin{equation}\label{eqn:elastMotionz}
\varrho \ddot{u}_z = \frac{\partial}{\partial x_j} c_{zjkl}^E
\frac{\partial u_l}{\partial x_k} + \frac{\partial}{\partial x}
e_{15} \frac{\partial \phi}{\partial x} + \frac{\partial}{\partial
y} e_{24} \frac{\partial \phi}{\partial y},
\end{equation}

\begin{eqnarray}\label{eqn:PiezoPoisson}
\nabla \cdot \epsilon^S_{s} \nabla \phi = \frac{\partial}{\partial
x} e_{15} \frac{\partial u_z}{\partial x} +
\frac{\partial}{\partial y} e_{24} \frac{\partial u_z}{\partial y}
+ \frac{\partial}{\partial z} e_{31} \frac{\partial u_x}{\partial
x} + \frac{\partial}{\partial x} e_{15} \frac{\partial
u_x}{\partial z} + \frac{\partial}{\partial z} e_{32}
\frac{\partial u_y}{\partial y} + \frac{\partial}{\partial y}
e_{24} \frac{\partial u_y}{\partial z}\ep
\end{eqnarray}
\end{widetext}

In Refs.~(\onlinecite{AizinGumbs1}) and (\onlinecite{AizinGumbs2})
analytical solutions of these equations were obtained, under the
assumption that the $\phi$ term in
Eqs.~\refeq{eqn:elastMotionx}--\refeq{eqn:elastMotionz} can be
ignored. This means that the mechanical motion is decoupled from
the electric field, but an electric field is induced from the
mechanical motion.  The reliability of such substantial
simplifications is limited to cases where the external potential
$\phi$ is small. Also, the obtained solutions are for
two-dimensional cases only and are therefore of minor interest
when studying the effect of surface gates. Solving the fully
coupled system of PDEs demands numerical techniques like the one
described in the next section.

\begin{widetext}
\section{Finite element formulation}
\label{sec:FEformulation}

We shall use the finite element method
in space and the finite difference method in time. The reasons for applying
the finite element method are the need for handling geometrically
complicated domains and the fact that the method works very well
for elasticity problems without coupling to $\phi$ as well as for
the Poisson equation for $\phi$.

We point out that the finite element formulation presented in this
paper is independent of the choice of element shape, as this would
be problem dependent. The elasticity part of the problem has three
degrees of freedom - one for each component of the displacement
vector. The electrical part has one degree of freedom for the
electric potential.

\subsection{The Poisson equation with
piezoelectric coupling} \label{sec:FEM}

Before continuing further, it is convenient to introduce a
superscript $\ell$ to denote the time level, for example,
$\phi^\ell$ is $\phi$ at time level $\ell$. The electrostatic
potential $\phi^\ell$ due to mechanical displacements $u_i^\ell$
in the piezoelectric material concerned is given by
Eq.~(\ref{eqn:PiezoPoisson}). The finite element formulation of
such a Poisson equation is well covered in the literature
\cite{Mohan,Iserles,DpBook2,Zienkiewicz}. The basic idea is to
approximate $\phi^\ell$ by a linear combination of basis functions
$N_i$, $\phi^\ell \approx \hat\phi^\ell = \sum_{j=1}^n
N_j\phi_j^\ell$, insert ${\hat\phi}^\ell$ in the Poisson equation,
and demand the residual to be orthogonal to the space spanned by
$\{ N_1,\ldots,N_n\}$. The Laplace term is integrated by parts.

The only non-trivial aspect of the formulation in the present
setting is the right-hand side, where the second-order derivatives
of the elastic field demand integration by parts. The two first
terms on the right-hand side of Eq.~(\ref{eqn:PiezoPoisson}) give
rise to an integral, which is straightforwardly integrated as

%\begin{widetext}
\begin{eqnarray}\label{eqn:PoissonFEM2}
- \int_{\Omega} N_i [\frac{\partial}{\partial x} e_{15}
\frac{\partial u_z^\ell}{\partial x}
+ \frac{\partial}{\partial y} e_{24} \frac{\partial u_z^\ell}{\partial y} ]d\Omega \nonumber \\
\nonumber \\ = \int_{\Omega} [e_{15} \frac{\partial N_i}{\partial
x} \frac{\partial u_z^\ell}{\partial x} + e_{24} \frac{\partial
N_i}{\partial y} \frac{\partial u_z^\ell}{\partial y} ]d\Omega  -
\oint_{\partial\Omega} N_i[ e_{15} \frac{\partial
u_z^\ell}{\partial x}n_x + e_{24} \frac{\partial
u_z^\ell}{\partial y}n_y] d\Gamma \ep
\end{eqnarray}
%\end{widetext}

The terms in Eq.~(\ref{eqn:PiezoPoisson}), containing mixed
derivatives, can be integrated by parts using a special form of
Green's Theorem,

%\begin{widetext}
\begin{equation}\label{eqn:green2}
    - \int_\Omega \phi \frac{\partial \psi}{\partial z} dx dy dz =
    \int_\Omega \frac{\partial \phi}{\partial z} \psi dx dy dz -
    \oint_\Gamma \phi \psi n_z d\Gamma \ep
\end{equation}
%\end{widetext}

The terms give rise to the integrals

%\begin{widetext}
\begin{eqnarray}
- \int_{\Omega} N_i [\frac{\partial}{\partial z} e_{31}
\frac{\partial u_x^\ell}{\partial x} + \frac{\partial}{\partial x}
e_{15} \frac{\partial u_x^\ell}{\partial z} +
\frac{\partial}{\partial z} e_{32} \frac{\partial
u_y^\ell}{\partial y} + \frac{\partial}{\partial y}
e_{24} \frac{\partial u_y^\ell}{\partial z} ]d\Omega \\ \nonumber \\
\nonumber = \int_\Omega [e_{31} \frac{\partial N_i}{\partial z}
\frac{\partial u_x^\ell}{\partial x} + e_{15} \frac{\partial
N_i}{\partial x} \frac{\partial u_x^\ell}{\partial z} +
e_{32}\frac{\partial N_i}{\partial z} \frac{\partial
u_y^\ell}{\partial y} + e_{24}\frac{\partial
N_i}{\partial y} \frac{\partial u_y^\ell}{\partial z} ]d \Omega \\ \nonumber \\
\nonumber - \oint_{\partial \Omega} N_i[e_{31} \frac{\partial
u_x^\ell}{\partial x} n_z + e_{15} \frac{\partial
u_x^\ell}{\partial z} n_x + e_{32} \frac{\partial
u_y^\ell}{\partial y} n_z + e_{24} \frac{\partial
u_y^\ell}{\partial z} n_y]d\Gamma \ep \label{eqn:PoissonFEM3}
\end{eqnarray}
%\end{widetext}

The finite element method applied to the equation for $\phi$
transforms the PDE problem to a linear system of equations,

\begin{equation}\label{eqn:femPoisson}
    \P \phivec^\ell = \f^\ell,
\end{equation}

\noindent where the `stiffness' matrix $\P$ has its $(i,j)$
element given by

\begin{equation}\label{eqn:femK}
    P_{ij} =  \int_{\Omega} \epsilon^S_s \left[
     \frac{\partial N_i}{\partial x} \frac{\partial N_j}{\partial x}
    +
     \frac{\partial N_i}{\partial y} \frac{\partial N_j}{\partial y}
    +
     \frac{\partial N_i}{\partial z} \frac{\partial N_j}{\partial
     z}\right]
     d\Omega\ \ep
\end{equation}

The $\phivec^\ell$ vector in Eq.~(\ref{eqn:femPoisson}) contains
the values of $\phi^\ell$ at the nodal points. The $i$-th
component $f_i^\ell$ of the right-hand side vector $\f^\ell$ can
be written as

\begin{eqnarray}\label{eqn:femF}
f_i^\ell =  \oint_{\partial \Omega} \epsilon^S_s N_i
\frac{\partial \phi^\ell}{\partial n}  d\Gamma - \oint_{\partial
\Omega} N_i r d\Gamma + \int_{\Omega} s d\Omega,
\end{eqnarray}
with
%\begin{widetext}
\begin{eqnarray*}\label{eqn:femF2}
r &=&  e_{31}\frac{\partial u_x^\ell}{\partial x} n_z  + e_{15}
\frac{\partial u_x^\ell}{\partial z} n_x + e_{32} \frac{\partial
u_y^\ell}{\partial y}n_z + e_{24} \frac{\partial
u_y^\ell}{\partial z}n_y +  e_{15} \frac{\partial
u_z^\ell}{\partial x} n_x + e_{24} \frac{\partial
u_z^\ell}{\partial y} n_y, \\
s&=&  e_{31} \frac{\partial N_i}{\partial z} \frac{\partial
u_x^\ell}{\partial x} +  e_{15} \frac{\partial N_i}{\partial x}
\frac{\partial u_x^\ell}{\partial z} +  e_{32} \frac{\partial
N_i}{\partial z} \frac{\partial u_y^\ell}{\partial y} + e_{24}
\frac{\partial N_i}{\partial y} \frac{\partial u_y^\ell}{\partial
z} + e_{15} \frac{\partial N_i}{\partial x}\frac{\partial
u_z^\ell} {\partial x} + e_{24} \frac{\partial N_i}{\partial y}
\frac{\partial u_z^\ell}{\partial y} \ep
\end{eqnarray*}
%\end{widetext}
When the displacement field entering $f_i^\ell$ is known,
any standard Poisson solver can be used to compute $\phivec^l$.

\subsection{The elasticity problem with electric field loading}
\label{NumericalFormulationElasticity}

For the finite element formulation of Eq.~(\ref{eqn:dispField1})
it is easier to first start with Eq.~(\ref{eqn:eqnMotion}) and
insert the constitutive law given by Eq.~(\ref{eqn:Tij1}) and the
strain-displacement relation given by Eq.~(\ref{eqn:Sij}) in the
finite element integrals. The time derivative in
Eq.~(\ref{eqn:dispField1}) can be approximated by a second-order
accurate finite difference. Sampling Eq.~(\ref{eqn:dispField1}) at
time level $\ell$ then yields

\beq \varrho { u^{\ell -1}_i -2u^\ell_i +u^{\ell+1}_i\over\Delta
t^2} = {\partial {\sigma_{ij}^\ell}\over\partial x_j} + \varrho
b_i^\ell , \label{tEL:eq1} \eeq

\noindent where $\Delta t$ is the time step length and quantities
with the $\ell$ superscript are functions of space only.
%Considering quantities at time level $\ell$ and $\ell -1$ as
%already computed, the only unknown in (\ref{tEL:eq1}) is
%$u^{\ell+1}_i$.
This time discretization introduces an operator splitting such
that the originally coupled governing equations can be solved in
sequence. No accuracy is lost by this operator splitting beyond
that implied by the finite difference itself in
Eq.~\refeq{tEL:eq1}. More precisely, $\sigma_{ij}^\ell$ contains
$\phi^\ell$, which is known, such that we can easily solve for
$u^{\ell+1}_i$ using old values of $\phi$. On the next time level
($\ell+1$), we can find $\phi^{\ell+1}$ using the recently
computed $u^{\ell+1}_i$.

The main motivation for the explicit time differencing in
Eq.~\refeq{tEL:eq1} is numerical efficiency: (i) we decouple the
equations, which simplifies the numerics and the implementation,
and (ii) there is no need to solve large sparse linear systems of
equations in the elasticity part of the problem (if the mass
matrix is lumped).  Nevertheless, the time difference
Eq.~\refeq{tEL:eq1} leads to a conditionally stable scheme, where
$\Delta t$ must be of the order of the smallest element size. In
wave propagation problems, uniform high resolution is frequently
needed in space and time, typically compatible with the stability
restriction, which makes such explicit schemes appropriate. On the
other hand, in applications where adaptive grids with great
variation in element size are needed, one may benefit from
implicit schemes (for example, of Newmark type)
\cite{Zienkiewicz}.

The finite element formulation of Eq.~(\ref{tEL:eq1}) is easiest
to express if we switch to a typical ``finite element
engineering'' notation, especially when we deal with anisotropic
media. The stress and strain tensors are expressed as vectors,
\beqan \tds &=& (\sigma_{xx},\sigma_{yy},\sigma_{zz},
\sigma_{yz},\sigma_{zx},\sigma_{yx})^T, \\
\tde &=& (\varepsilon_{xx}, \varepsilon_{yy}, \varepsilon_{zz},
2\varepsilon_{yz}, 2\varepsilon_{zx}, 2\varepsilon_{yx})^T \ep
\eeqan The constitutive law can then be written as \beq \tds^\ell
= \D\tde^\ell + \p^\ell,\label{EL:eq14} \eeq where the $\p^\ell$
term represents the loading from the electric field, and $\D$ is a
symmetric $6\times 6$ matrix of the elasticity coefficients
(previously denoted by $c_{ijkl}^E$):

%\begin{widetext}
\begin{equation}\label{eqn:cIJ}
   \D=\left[%
\begin{array}{cccccc}
  c_{11} & c_{12} & c_{13} & c_{14} & c_{15} & c_{16}
  \\ \\
   & c_{12} & c_{13} & c_{14} & c_{15} & c_{16}
  \\ \\
   &  & c_{13} & c_{14} & c_{15} & c_{16}
  \\ \\
 &  &  & c_{14} & c_{15} & c_{16}
  \\ \\
   &  &  &  & c_{15} & c_{16}
  \\ \\
   &  &  &  &  & c_{16}
\end{array}%
\right] \ep
\end{equation}

The displacement field at a time level $\ell$ is approximated
according to

\[\ u_i^\ell \ \approx\hat\u^\ell = \sum_{j=1}^n N_j\u_j^\ell,\]
where $\u_j^\ell$ is the value of the displacement field at node $j$ at
time level $\ell$.
The strain-displacement relation then becomes
\beq \tde^\ell = \sum_{j=1}^n \B_j\u_j^\ell ,
\quad
\B_i =\left[\begin{array}{ccc}
N_{i,x} & 0 & 0\\
0 & N_{i,y} & 0 \\
0 & 0 & N_{i,z}\\
0& N_{i,z} & N_{i,y}\\
N_{i,z} & 0 & N_{i,x} \\
N_{i,y} & N_{i,x} & 0
\end{array}\right] \ep
\label{EL:eBu}
\eeq

\noindent The comma notation here denotes partial derivative:
$N_{i,x}\equiv \partial N_i/\partial x$.

A finite element formulation of Eq.~(\ref{tEL:eq1}) using the
aforementioned notation becomes (see Ref.~\onlinecite{DpBook2} for
a detailed derivation without the $\phi$ term)

%\begin{widetext}
\beqa \int\limits_\Omega \varrho N_i\sum_{j=1}^n N_j (\u^{\ell
-1}_j -2\u^\ell_j + \u^{\ell+1}_j)d\Omega + \Delta
t^2\int\limits_{\Omega}\B_i^T\tds^\ell d\Omega &=& \Delta
t^2\int\limits_{\Omega} \varrho \b N_i d\Omega + \Delta
t^2\int\limits_{\partial\Omega} \bft^\ell N_i d\Gamma \ep
\label{EL:BsdO} \eeqa Here, $\bft^\ell$ is the traction on the
boundary, arising from integrating the divergence of the stress in
Eq.~(\ref{tEL:eq1}) by parts. Inserting the constitutive law in
the term $\int\limits_{\Omega} \B_i^T\tds d\Omega$ yields
%\end{widetext}
\[
\sum_{j=1}^n \left(\int\limits_{\Omega} \B_i^T\D\B_j d\Omega \right)\u_j^\ell +
\int\limits_{\Omega} \B_i^T\p^\ell d\Omega \ep
\]

The final discrete equations arising from the equation of motion
can be written as

\beq \u^{\ell +1} = 2\u^{\ell} -\u^{\ell -1} + \Delta
t^2\tilde\M^{-1}(-\K\u^\ell + \modvec{\beta}^\ell +
\modvec{\Phi}^\ell) \label{waves1}  , \eeq

\noindent where $\tilde\M^{-1}$ is an efficient inverse of the
mass matrix $\M$ resulting from the $\int_\Omega\varrho
N_iN_jd\Omega$ integral. We construct $\tilde\M^{-1}$ as the
inverse of the lumped mass matrix (using the row-sum technique to
lump the matrix). The matrix $\K$ stems from the
standard ``stiffness'' term $\int_{\Omega}\B_i^T\D\B_j d\Omega$
representing anisotropic elasticity, $\modvec{\beta}^\ell$ is the
effect of body forces and surface tractions, and
$\modvec{\Phi}^\ell$ is the contribution from the electric field.
The $i$-th block (arising
from node $i$) in $\modvec{\Phi}^\ell$ takes the form
$\modvec{\Phi}^\ell_i = \int_\Omega\B_i^T\p^\ell$, which from
Eqs.~(\ref{eqn:elastMotionx}) to (\ref{eqn:elastMotionz}) results
in
%\begin{widetext}
\[
\modvec{\Phi}^\ell_i = \left[\begin{array}{c} \int_{\Omega} N_i[
\frac{\partial}{\partial x} e_{31} \frac{\partial
\phi^\ell}{\partial z} + \frac{\partial}{\partial z} e_{15}
\frac{\partial \phi^\ell}{\partial x}]d \Omega \\ \\
-\int_{\Omega}  N_i [\frac{\partial}{\partial y} e_{32}
\frac{\partial \phi^\ell}{\partial z} + \frac{\partial}{\partial
z} e_{24}\frac{\partial \phi^\ell}{\partial y} ]d \Omega  \\ \\
\int_{\Omega} N_i [\frac{\partial}{\partial x}e_{15}\frac{\partial
\phi^\ell}{\partial x} + \frac{\partial}{\partial
y}e_{24}\frac{\partial \phi^\ell}{\partial y}] d \Omega
\end{array}\right]
\]
\\ \\
\beq  = \left[\begin{array}{c} - \int_{\Omega} [e_{31}
\frac{\partial N_i}{\partial x}  \frac{\partial
\phi^\ell}{\partial z} + e_{15} \frac{\partial N_i}{\partial z}
\frac{\partial \phi^\ell}{\partial x} ]d \Omega + \int_{\partial
\Omega} N_i [e_{31} \frac{\partial \phi^\ell}{\partial z} n_x +
e_{15} \frac{\partial \phi^\ell}{\partial x}
n_z] d \Gamma \\ \\
 \int_{\Omega} [e_{32} \frac{\partial N_i}{\partial y}
\frac{\partial \phi^\ell}{\partial z} + e_{24} \frac{\partial
N_i}{\partial z} \frac{\partial \phi^\ell}{\partial y} ]d \Omega -
 \int_{\partial \Omega} N_i [e_{32}
\frac{\partial \phi^\ell}{\partial z}  n_y + e_{24}
\frac{\partial \phi^\ell}{\partial y}  n_z]d \Gamma \\ \\
-\int_{\Omega} [e_{15} \frac{\partial N_i}{\partial x}
\frac{\partial \phi^\ell}{\partial x} + e_{24} \frac{\partial
N_i}{\partial y} \frac{\partial \phi^\ell}{\partial y} ]d\Omega +
\int_{\partial \Omega} N_i [e_{15} \frac{\partial
\phi^\ell}{\partial x} n_x + e_{24}\frac{\partial
\phi^\ell}{\partial y} n_y]d \Gamma
\end{array}\right] \ep
\label{eqn:Phiterm} \eeq%Elasticity solvers which perform the
%computation given by (\ref{waves1}) (without the $\modvec{\Phi}$
%term) are widely available.
%\end{widetext}

Our governing equations require initial conditions. Let us assume
that the elastic body is at rest such that ${\partial
u_i}/{\partial t} = 0$. The external electric field is then turned
on. After the initial transients the displacement field have faded
out, and we have a stationary initial state of our system, modeled
by the equations \beqa
    \P \phivec^0 &=& \f^0 (\u^0),\label{eqn:PoissonTeq0}\\
     \K\u^0 &=& \modvec{\beta}^0 + \modvec{\Phi ^0(\phi^0)}
\label{eqn:statElastTeq0} \ep \eeqa These two coupled equations
are solved by an iterative Gauss-Seidel-like technique, i.e., the
equations are solved one at a time, using the most recent
approximation of the other field in the right-hand side term. Each
linear system is solved by a MILU preconditioned conjugate
gradient method \cite{Langtangen:89b}. The solution of
Eqs.~\refeq{eqn:PoissonTeq0} and \refeq{eqn:statElastTeq0}, along
with the assumption of stationarity, $\partial\u /\partial t = 0$,
comprise the initial condition.

As boundary conditions, we either have prescribed traction components
or prescribed displacement components, along with prescribed electric
potential or prescribed normal component of the electric field.
The displacement equation needs three boundary conditions at each point
at the boundary, while the equation for $\phi$ needs one condition at
each point.

From $\partial\u /\partial t = 0$ at $t=0$ it follows by a
second-order difference approximation that $\u^1 = \u^{-1}$. From
Eqs.~\refeq{waves1} and \refeq{eqn:statElastTeq0} it follows that
$\u^1=\u^0=\u^{-1}$. The loads removed at $t=0^+$ will first come
into play at the second time level.

The computational algorithm can now be formulated as follows:

\begin{quote}
\begin{tabbing}
\hspace*{0.5cm}\= \hspace{0.5cm} \= \hspace{0.5cm} \=
\hspace{0.5cm} \= \hspace{0.5cm} \= \kill

$\ell = 0$ (time level counter) \\
$k = 0$ (iteration counter) \\

\mbox{while} $\varepsilon < \varepsilon_{crit}$\\
\> solve Eq.~(\ref{eqn:PoissonTeq0}) w.r.t ${\phivec}^{0,k}$ \\
\> solve Eq.~(\ref{eqn:statElastTeq0}) w.r.t ${\u}^{0,k}$ \\
\> compute $\varepsilon = \parallel {\u}^{0,k} -
{\u}^{0,k-1}\parallel + \parallel \phivec^{0,k} - \phivec^{0,k-1} \parallel$ \\
\>$k \leftarrow k + 1$
\mbox{end while} \\
$\phivec^{1} = \phivec^{0}$, $\u^{1} = \u^{0}$ \\
for $\ell = 1,2,3,\ldots$ until end of simulation\\
\> solve Eq.~(\ref{waves1}) w.r.t ${\u}^{\ell+1}$ \\
\> solve Eq.~(\ref{eqn:femPoisson}) w.r.t ${\phivec}^{\ell+1}$ \\
\end{tabbing}
\end{quote}

\noindent {\bf Comments on Stability.} The stability criterion of
the explicit finite difference scheme in time, when decoupled from
the electric field problem, requires $\Delta t \leq
2/\omega_{\max}$ \cite{Cook}, where $\omega_{\max}$ is the highest
natural frequency of the vibrating system. One may find
$\omega_{\max}$ as the square root of the largest eigenvalue of
the problem $\K - \lambda \M =0$. An approximate bound on
$\omega_{\max}$ can be estimated from a relation $\omega_{\max}=
2c/h_{\rm eff}$, where $c$ is the speed of elastic waves and
$h_{\rm eff}$ is the smallest effective element length. The
stability criterion now reduces to the common Courant, Friedrichs,
and Lewy (CFL) condition: \beq \Delta t \leq \alpha h_{\rm eff}/c
\ep \eeq Here, $\alpha$ is a factor to be adjusted since the
$h_{\rm eff}$ parameter is normally roughly computed from element
sizes or application of Gerschgorin's theorem applied to the
underlying eigenvalue problem. For wave problems the CFL condition
is frequently not particularly restrictive since the length and
period of a wave are usually proportional, leading to a natural
choice of $\Delta t/h=\hbox{const}$.

\section{An optimised finite element formulation}
\label{sec:optimisation}

We can speed up the numerical
computation by replacing the complete assembly of the right-hand
side vectors in Eqs.~(\ref{eqn:femPoisson}) and (\ref{waves1}) by
a matrix-vector product. For example, consider the contribution to
the vector $\bf{\f}$ in Eq.~(\ref{eqn:femPoisson}) due to the
coupling from the mechanical motion (omitting the superscript
$\ell$ for the remainder of this section):

%\begin{widetext}
\beq\label{eqn:optimisation} \int_\Omega e_{31} \frac{\partial
N_i}{\partial z} \frac{\partial u_x}{\partial x} +
e_{15}\frac{\partial N_i}{\partial x} \frac{\partial u_x}{\partial
z} + e_{32} \frac{\partial N_i}{\partial z} \frac{\partial
u_y}{\partial y} + e_{24} \frac{\partial N_i}{\partial y}
\frac{\partial u_y}{\partial z} +  e_{15} \frac{\partial
N_i}{\partial x}\frac{\partial u_z} {\partial x} + e_{24}
\frac{\partial N_i}{\partial y} \frac{\partial u_z}{\partial y} d
\Omega. \eeq

\noindent Inserting the finite element expansion $u_x\approx
\sum_{j=1}^n u_j^{(x)}N_j$, and similar expansions for $u_y$ and
$u_z$, the expression in Eq.~\refeq{eqn:optimisation} results in

\begin{eqnarray}\label{eqn:optimisation2}
\sum_{j=1}^n \int_\Omega e_{31} \frac{\partial N_i}{\partial z}
\frac{\partial N_j}{\partial x} u_j^{(x)} + e_{15} \frac{\partial
N_i}{\partial x} \frac{\partial N_j}{\partial z} u_j^{(x)} +
e_{32} \frac{\partial N_i}{\partial z} \frac{\partial
N_j}{\partial y} u_j^{(y)}  \\ \nonumber \\
\nonumber  + e_{24} \frac{\partial N_i}{\partial y}\frac{\partial
N_j}{\partial z} u_j^{(y)} + e_{15} \frac{\partial N_i}{\partial
x}\frac{\partial N_j} {\partial x} u_j^{(z)} + e_{24}
\frac{\partial N_i}{\partial y} \frac{\partial N_j}{\partial y}
u_j^{(z)}  d \Omega.
\end{eqnarray}

This expression can be written as a matrix vector product
$\L \u$,
where the matrix $\L$ consists of $n \times n$ blocks, each
block of size $1 \times 3$. For the coupling of node $i$ and $j$,
the $3 \times 1$ block looks like

\begin{eqnarray}\label{eqn:optimisation3}
e_{31} \frac{\partial N_i}{\partial z} \frac{\partial
N_j}{\partial x} + e_{15} \frac{\partial N_i}{\partial x}
\frac{\partial N_j}{\partial z} ,  e_{32} \frac{\partial
N_i}{\partial z} \frac{\partial N_j}{\partial y} + e_{24}
\frac{\partial N_i}{\partial y} \frac{\partial N_j}{\partial z},
e_{15} \frac{\partial N_i}{\partial x}\frac{\partial N_j}
{\partial x} + e_{24} \frac{\partial N_i}{\partial y}
\frac{\partial N_j}{\partial y},
\end{eqnarray}

The Poisson equation - Eq.~(\ref{eqn:femPoisson}) for $\phi$ can
now be written as
\begin{equation}\label{eqn:femPoissonOpt}
    \P \phivec^{\ell+1} = \L \u^{\ell+1}\ep
\end{equation}
At each time level we can hence avoid the costly finite element
assembly process, since $\P$ and $\L$ are constant in time and the
right-hand side of the linear system is obtained by an efficient
matrix-vector product. This may result in a significant speed-up
of the solver. For large number of unknowns, the speed up is
experienced only if we use a method of complexity or order $n$,
like multigrid, to solve Eq.~\refeq{eqn:femPoissonOpt}.

Equation~(\ref{waves1}) is already on a favorable matrix-vector
algebra form, except for the coupling term $\modvec{\Phi}^\ell$.
Examining Eq.~\refeq{eqn:Phiterm}, we realize that this matrix can
be written as $\modvec{\Phi}^{\ell}=\Q\phivec^{\ell}$, where $\Q$
is a time-independent $n\times n$ matrix of $3\times 1$ blocks.
The contribution to block $i$ in $\modvec{\Phi}^{\ell}$ is made of
the coupling $\Q_{ij}$ between node $i$ and $j$ and the block $j$
in $\phivec^{\ell}$: $\modvec{\Phi}_i^{\ell} =
\Q_{ij}\phivec_j^{\ell}$, where

\[ \Q_{ij} =
 \left[\begin{array}{c} - \int_{\Omega} e_{31} \frac{\partial
N_i}{\partial x}  \frac{\partial N_j} {\partial z} + e_{15}
\frac{\partial N_i}{\partial z}  \frac{\partial N_j} {\partial x}
d \Omega + \int_{\partial \Omega} N_i [e_{31}  \frac{\partial
N_j}{\partial z} n_x + e_{15}  \frac{\partial N_j}{\partial x}
n_z ] d \Gamma \\ \\
 \int_{\Omega} e_{32} \frac{\partial N_i}{\partial y}
\frac{\partial N_j}{\partial z} + e_{24} \frac{\partial
N_i}{\partial z} \frac{\partial N_j}{\partial y} d \Omega -
\int_{\partial \Omega} N_i [ e_{32} \frac{\partial N_j}{\partial
z} n_y + e_{24}
\frac{\partial N_j}{\partial y}  n_z ] d \Gamma \\ \\
-\int_{\Omega} e_{15} \frac{\partial N_i}{\partial x}
\frac{\partial N_j}{\partial x} + e_{24} \frac{\partial
N_i}{\partial y} \frac{\partial N_j}{\partial y} d\Omega +
\int_{\partial \Omega}
 N_i [e_{15} \frac{\partial N_j}{\partial x}  n_x + e_{24} \frac{\partial
N_j}{\partial y} n_y]d \Gamma
\end{array}\right]\ep
\]

Now the $\modvec{\Phi}^\ell$ term in Eq.~(\ref{waves1}) can be
computed as a matrix-vector product $\Q\phivec^\ell$, and no
assembly process is required to construct Eq.~(\ref{waves1}) at
each time level. In our algorithm we must run assembly processes
at $t=0$ to construct the matrices $\tilde\M^{-1}$, $\K$, $\Q$,
$\L$, and $\P$.
\end{widetext}

\section{Object-oriented implementation} \label{sec:Implementation}

A significant trend in modern software development is to formulate
numerical algorithms such that reliable and well-tested software
components can be combined together to form a new simulator. Our
numerical approach was in particular inspired by such an approach.
With the time discretization we were able to split the coupled
$u_i$--$\phi$ system such that at each time level we first solve
for a new displacement field ($u_i$) and then we solve for the
corresponding electric potential ($\phi$). In each of the two
equations, the effect of the other only enters through a
right-hand side ``forcing'' term. This allows us, in principle, to
reuse a solver for elastic vibrations and a Poisson solver, as
long as these solvers can implement a new right-hand side term
that couples to another solver. From a principal point of view,
this idea is simple and attractive. However, the implementation of
the idea in practice may be less feasible if the design of the
underlying solvers is not sufficiently flexible.

To allow for building the compound $u_i$--$\phi$ solver from
separate $u_i$ and $\phi$ solvers we propose to apply principles
from object-oriented programming. The time-dependent elasticity
solver and the Poisson solver are realized as two independent
objects, implemented via classes in C++. Actually, we have reused
the \code{Poisson2} class from Ref.~\onlinecite{DpBook2} as the
Poisson solver, and the \code{ElasticVib1} class from
Ref.~\onlinecite{DpBook2} as the elastic vibration solver. The
latter is a subclass of \code{Elasticity2}, a pure quasi
time-dependent elasticity solver (without the acceleration term in
the momentum equation). These classes are implemented using
building blocks from the Diffpack library \cite{DpBook2}.

The \code{Poisson2} and \code{ElasticVib1} classes have no
knowledge of each other. The only common feature is their design.
This design is crucial for reuse of the classes to solve the
coupled problem, but the design approach suggested for Diffpack
solvers \cite{DpBook2,TOMS2001,Munthe:00} has proven to be
successful in this respect. Diffpack solvers are realized as
classes containing objects for grids, scalar/vector fields, linear
systems, etc. The coefficients in a PDE are evaluated through
virtual functions. When the solver is stand-alone, these virtual
functions contain mathematical expressions or measured data, but
when solvers are combined, the virtual functions are reimplemented
in subclasses and connected to data in other solvers. This
principle reflects the underlying mathematics: the coefficients in
a single PDE are considered known, but in a system of PDEs, the
coefficients typically couple to unknown quantities governed by
the PDEs. The present coupled system is an example where the
right-hand side in the Poisson equation couples to the primary
unknown in the elastic vibration solver, while a right-hand side
term in the elastic vibration solver couples to the primary
unknown in the Poisson equation. When the solvers are used
independently there are no such couplings.

\begin{figure}[!h]
\begin{center}
\rotatebox{0}{\scalebox{0.4}{\includegraphics*[1.0in,1.5in][7.5in,10.0in]{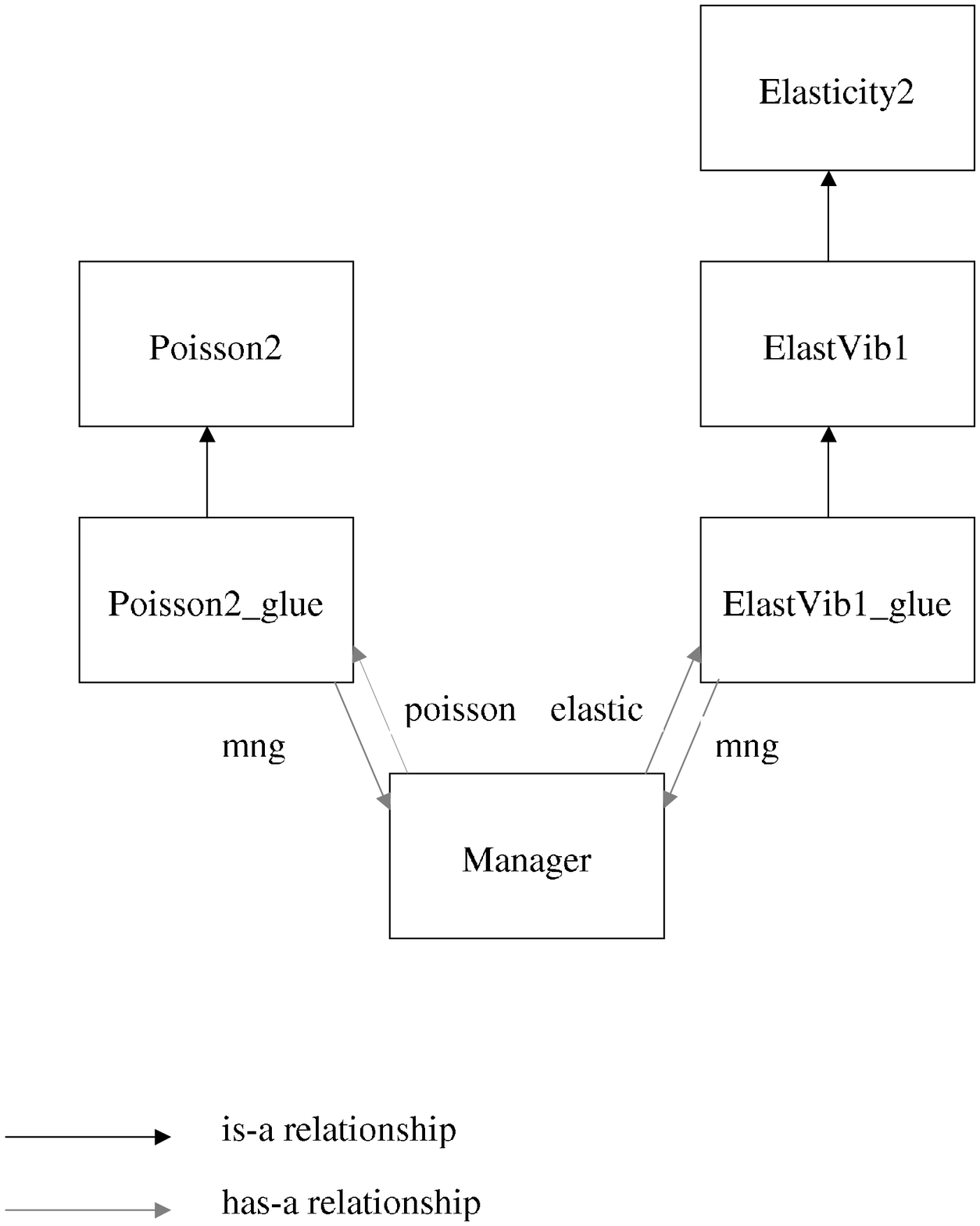}}}
\end{center}
\caption{Relationships between classes. The base classes,
\code{Poisson2} and \code{Elasticity2} perform the solution of the
standard time-independent Poisson and elasticity equations,
respectively. \code{ElasticVib1} is derived from \code{Elastcity2}
and solves for time-dependent elastic motion. The classes
\code{Poisson2\_glue} and \code{ElasticVib1\_glue} implement the
coupling between the electrostatic and mechanical equations, and
these classes are controlled by the \code{Manager} class.}
\label{fig:classes}
\end{figure}

Figure~\ref{fig:classes} shows an outline of the class design for
the compound solver. From class \code{Poisson2} we derive a
subclass \code{Poisson2\_glue}, which reimplements the virtual
function in \code{Poisson2} for evaluating the right-hand side in
that equation. In this new function we need to compute an
expression involving the displacement field available in the
\code{ElasticVib1} class. A manager class, simply called
\code{Manager}, holds pointers to all classes for the individual
PDEs in the system of PDEs, and each PDE class holds a pointer to
the manager class, thus enabling a two-way communication. With
these pointers, we can connect data from any other solver class to
the \code{Poisson2\_glue} class. In the virtual function
evaluating the right-hand side we can typically call
\code{mng->elastic->u} to get a vector field object for $u_i$ that
we can evaluate at the current integration point. This object has
support for evaluating derivatives of the field as well.

Similarly, we derive a subclass \code{ElasticVib1\_glue} of
\code{ElasticVib1}, add a pointer to the manager class, and
override the virtual function for evaluating the right-hand side.
Now we need to compute expressions involving $\phi$, but this is
easily accomplished by the pointers, e.g., \code{mng->poisson->u}
if \code{u} is the name of the unknown scalar field in the
\code{Poisson2} solver.

The ``glue'' classes \code{Poisson2\_glue} and
\code{ElasticVib1\_glue} are very small compared to the real
solver classes they inherit from. A ``glue'' class typically needs
about a page of code unless it adds additional computations.
The manager class is a bit more comprehensive since it implements
the overall solution algorithm, i.e., the time stepping and the
calls to the independent solvers.

The benefit from using object-oriented programming in the way we
have outlined is that the original well-tested Poisson and elastic
vibration solvers can be reused in a system of PDEs without any
modifications. The original solver classes reside in separate
files and are hence not subject to any side effects from editing
the source code. The ``glue'' class is also in a separate file and
enables the original solver to speak to a manager in charge of
solving a compound system of PDEs. The advantages are clear:
reliability is increased by reusing well-tested solvers, and the
compound solver is modularized. Our experience is that this design
reduces the development time significantly. Especially the
debugging phase is greatly simplified.
At any time, the underlying solvers can
be trivially pulled out of the compound system and verified
independently.

\section{Verification} \label{sec:verification}
%\subsection{Analytic solutions}

%\begin{figure}[!h]
%\begin{center}
%\rotatebox{0}{\scalebox{0.5}{\includegraphics*[0.5in,7.0in][7.5in,11.5in]{beamGr.ps}}}
%\end{center}
%\caption{The beam used for testing the solver. The scale shows the
%error in the potential.} \label{fig:beam}
%\end{figure}

Before showing numerical results for a physically relevant
application, we report on the estimated numerical accuracy of the
code, thereby establishing evidence for the correctness of the
implementation. To investigate numerical errors, we compare numerical
results with an exact solution. The exact solution is based on the
assumption of a displacement field in $z$ direction only, depending on
$x$ and $t$ only. Physically, a normal traction is applied at all
$x=\mbox{const}$ and $y=\mbox{const}$ boundaries to avoid
displacements in the $x$ and $y$ directions, but in the code we
implement this situation by essential boundary conditions
$u_x=u_y=0$. We also assume that $\phi =\phi(x,t)$ and that the
material is isotropic. The governing equations then reduce to
\begin{eqnarray}\label{eqn:elastMotionz1b}
\varrho\ddot{u}_z &=&  \mu\frac{\partial^2 u_z}{\partial x^2} +
e_{15}\frac{\partial^2\phi}{\partial x^2},\\
\epsilon_s^S\frac{\partial^2\phi}{\partial x^2} &=&
e_{15}\frac{\partial^2 u_z}{\partial x^2}\ep
\end{eqnarray}
These equations can be scaled to yield
\begin{eqnarray}\label{eqn:elastMotionz1}
\ddot{u}_z &=&  \frac{\partial^2 u_z}{\partial x^2} + \alpha
\frac{\partial^2\phi}{\partial x^2},\\
\frac{\partial^2\phi}{\partial x^2} &=&  \frac{\partial^2 u_z}{\partial x^2}\ep
\label{eqn:elastMotionz13}
\end{eqnarray}
The constant $\alpha$ is zero or unity corresponding to whether
the elasticity problem couples to the electric field problem or
not. (The three coefficients in the original system are scaled
away by choosing an appropriate time scale, $\phi$ scale, and
$u_i$ scale.) One possible solution of
Eqs.~\refeq{eqn:elastMotionz1}--\refeq{eqn:elastMotionz13} reads
\begin{eqnarray}\label{eqn:sepvar3}
    u_z(x,t) = \phi(x,t) = \cos (k\sqrt{1+\alpha} t) \sin (k x)\ep
\end{eqnarray}
In our tests we choose $\Omega$ as a three-dimensional
beam and $k = \frac{n \pi}{L}$, with $n$
being an integer and $L$ the length of the beam. The ends of the
beam are then fixed.

%Figure \ref{fig:beam} shows the error field $e=u_z-\hat u_z$,
%where $u_z$ is the exact solution and $\hat u_z$ is the computed
%finite element approximation, on the surface of the beam. The
%actual potential had an amplitude of 0.01 in this problem, and
%therefore according to the scale, the errors are less than 0.01
%percent in this particular run.

Our verification procedure consists of estimating the convergence
rate of the numerical method. A scalar error measure $e$ is
expected to behave like
\[ e = Ah^r + B\Delta t^s,\]
where $A$ and $B$ are constants independent of the element size
$h$ and the time step $\Delta t$.
From the involved approximations,
we expect $s=2$ and $r=1+q$, where $q$ is the order of the
polynomials in an element. In particular, for linear or trilinear
elements, $r=2$, and if $h$ and $\Delta t$ are chosen such that
$h/\Delta t = \hbox{const}$, we see that
$e\sim h^r$. From two successive experiments, $(h_{i-1},e_{i-1})$ and
$(h_i,e_i)$ we may estimate a convergence rate $r_i$ from
\[ r_i = {\ln e_i/e_{i-1}\over\ln h_i/h_{i-1}}\ep\]

We have investigated three error measures, given as different norms of the
error field $E=u_z -\hat u_z$, where $\hat u_z$ is the numerical solution
and $u_z$ is the exact solution:

\begin{widetext}
\[|| E ||_{L^1(\Omega)} = \int_{\Omega} |E| d \Omega ,
\quad
    || E ||_{L^2(\Omega)} = ( \int_{\Omega} E^2 d \Omega
    )^{\frac{1}{2}},\quad
    || E ||_{L^{\infty}(\Omega)} = \hbox{sup} (  |E(\textbf{x})|,\,\,
    \textbf{x}\in\Omega) \ep
\]
\end{widetext}
Table~I displays the values of the error norms and the associated
estimates of the convergence rates $r_i$. As the grid spacing $h$
and the time step $\Delta t$ are reduced, the numerical solution
converges to the exact solution with the expected rate of two.
This provides evidence for the correctness of the implementation
and indicates that our overall solution method, including the
operator splitting, is of second order in time and space. The
spatial convergence rate is expected to increase with the order of
the polynomials used in the elements.

\begin{table}[!h]\label{tab:NRsaw}
\centering
\begin{tabular}{|l|r|r|r|r|r|r|}
\hline
% after \\: \hline or \cline{col1-col2} \cline{col3-col4} ...
$h$  & $|| E ||_{L^1}$ & rates & $|| E ||_{L^2}$ & rates & $|| E ||_{L^{\infty}}$ & rates  \\
\hline
0.05        & $2.440\cdot 10^{-5}$ &       & $2.710\cdot 10^{-6}$ &       & $3.830\cdot 10^{-7}$ &     \\
0.1         & $9.760\cdot 10^{-5}$ & 1.999 & $1.084\cdot 10^{-6}$ & 2.000 & $1.533\cdot 10^{-6}$ &  2.000   \\
0.2         & $3.902\cdot 10^{-4}$ & 1.999 & $4.335\cdot 10^{-5}$ & 1.999 & $6.130\cdot 10^{-6}$ &  1.999   \\
0.25        & $6.096\cdot 10^{-4}$ & 1.999 & $6.771\cdot 10^{-5}$ & 1.999 & $9.574\cdot 10^{-6}$ &  1.999   \\
0.5         & $2.432\cdot 10^{-3}$ & 1.996 & $2.702\cdot 10^{-4}$ & 1.996 & $3.819\cdot 10^{-5}$ &  1.996   \\
\hline
\end{tabular}
\caption{Error norms and estimated convergence rates for waves in a
piezoelectric material, with $h = 10\Delta t$.}
\end{table}

\section{A surface acoustic wave application} \label{sec:Application}
In order to show that our suggested numerical model may be applied
to real-world phenomena, we apply it to a problem in quantum
electronics, specifically in acoustic charge transport. We
simulate a SAW propagating through a nano-scale substrate  of
GaAs. SAWs are particular solutions of
Eqs.~(\ref{eqn:elastMotionx}) - (\ref{eqn:PiezoPoisson}) such that
they propagate without decay on the surface of a material but
decay exponentially into the bulk. The solutions typically have
the form,

\begin{eqnarray}\label{eqn:SAW}
    u_i = \sum_j U_{i,j} e^{- k q_j z} e^{i(kx -\omega t)}, \\ \nonumber
    \phi = \sum_j \Phi_j e^{-k q_j z} e^{i(kx - \omega t)},
\end{eqnarray}

\noindent assuming $x$ is the direction of propagation, $z$ is the
direction into the bulk ($z\rightarrow-\infty$), and the decay
constants $q_j$, may be complex, allowing for oscillatory decay
(in an exponential envelope) into the bulk as is the case for GaAs
\cite{Simon}. The $U_{i,j}$ and $\Phi_j$ are constant amplitudes,
and $k$ is a wavenumber. It is a straightforward mathematical
procedure to determine the decay constants and wave velocity
\cite{Matthews}. However, we are interested in the more
complicated dynamics taking place when an obstacle in the form of
a charged metallic gate is placed on the surface.

In acoustic charge transport, the time-varying electric potential
accompanying the SAW is used to transport electrons which are
trapped in the minima of the waves. To perform further
manipulation on these electrons, for example, to remove some from
a minimum, external electric fields must be applied, and this is
achieved by applying voltages to metallic gates placed on the
surface. In general, the gates have different electrical and
mechanical properties to the bulk piezoelectric material and so we
may expect to see interesting effects if for example a SAW is
passed through the compound structure. Here, we perform three
simulations where we pass a SAW through a piece of GaAs. The first
will be a bare SAW, as a check to see we do excite the required
modes. The following two will have a metallic gate of contrasting
mechanical properties to GaAs with a static voltage applied on it.
In the first of these, we decouple the mechanical motion from the
electric field (while keeping the electric field coupled to the
mechanical motion). In the second we allow the full mutual
coupling between the electrical and mechanical fields to take
place. The dimensions of the gate are 600~nm $\times$ 400~nm
$\times$ 200~nm. To magnify the effect of the compound material
structure we use a fictitious material for the gate, similar in
crystal structure to GaAs, but with the elastic constants and mass
density changed by one order of magnitude. Moreover, we apply a
relatively large voltage of 1.5~V so that the coupling of the
mechanical displacement to the electric field is evident.

In the numerical simulations we have implemented 8-noded
hexahedral brick elements with $2 \times 2 \times 2$
Gauss-Legendre integration. The spacing along the direction of the
SAW propagation - $\Delta x$ is chosen to be $100$~nm. The time
integration parameter $\Delta t$ is $1.25 \times 10^{-2}$ ns.

\subsection{Experiments with surface acoustic waves}

In realistic SAW-based devices, bulk waves and SAWs are excited by
the application of a microwave signal to an interdigitated
transducer. The waves propagate outward from the source, with bulk
waves dissipating energy into the bulk and surface acoustic waves
travelling through the free surface. In acoustoelectric charge
transport experiments, the quantum devices are located several
thousand microns away from the source where one would expect to
see only surface acoustic waves. To simulate the entire system
would therefore require vast computational resources (the
wavelength of a SAW is one micron and approximately ten nodes
would be required across one wavelength for a satisfactory
description). Instead we describe a system of boundary conditions
which excite acoustic waves with SAWs dominating over the bulk
waves within a few microns of the source.

\subsubsection{Excitation of SAWs.}

We point out that to be consistent with the formulation presented
in the previous sections and hence acoustoelectric charge
transport experiments, in our simulations the SAW travels along
the crystal positive [011] axis, with the positive $z$ axis
aligned with the crystal positive [100] axis.

\begin{figure}[!h]
\epsfxsize=7.5cm \centerline{\epsffile{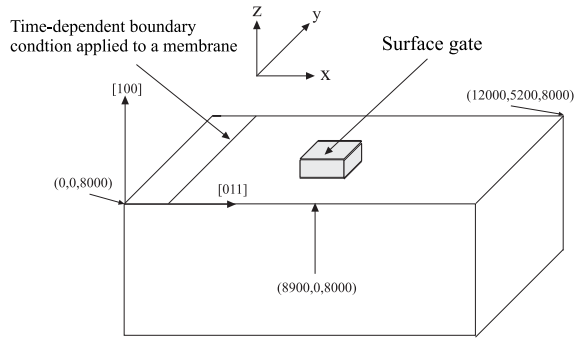}}

\caption{A schematic diagram of the experimental setup for
simulating SAWs. The gate is centered at ($8.9\cdot 10^{3}$~nm,
$2.6\cdot 10^{3}$~nm) } \label{fig:model}
\end{figure}

In order to excite the SAW modes, we apply a time-dependent
Dirichlet boundary condition to the $z$ component of the
displacement field on a small region on the grid. This can be
expressed as

%\begin{widetext}
\begin{eqnarray}
\label{eqn:DispBC}
    u_x = u_y = 0,\quad
    u_z = A \sin (2 \pi f t), \\ \nonumber  \qquad  x_0 \leq x \leq x_1, \quad
    z=8000~\text{nm} \ep
\end{eqnarray}
%\end{widetext}

The thickness of the membrane is defined as $x_1 - x_0 = 100$~nm.
To reduce diffraction effects from the surfaces - $y=0$~nm and
$y=5200$~nm we have forced $u_y = 0$ on those faces. The frequency
$f$ used is 2.7~GHz and the amplitude $A$ is chosen so that the
SAW amplitude is approximately 20~mV at a depth of 100~nm.
Traction free boundary conditions for the mechanical equations and
zero normal electrical displacements are implemented at the
surfaces. We found this system of boundary conditions to be a
particularly efficient means of generating coherent SAWs. The
setup is shown in Fig.~\ref{fig:model}.

\begin{figure}[!h]
\begin{center}
\scalebox{0.3}[0.22]{\includegraphics*[0.0in,0.5in][9in,12.0in]{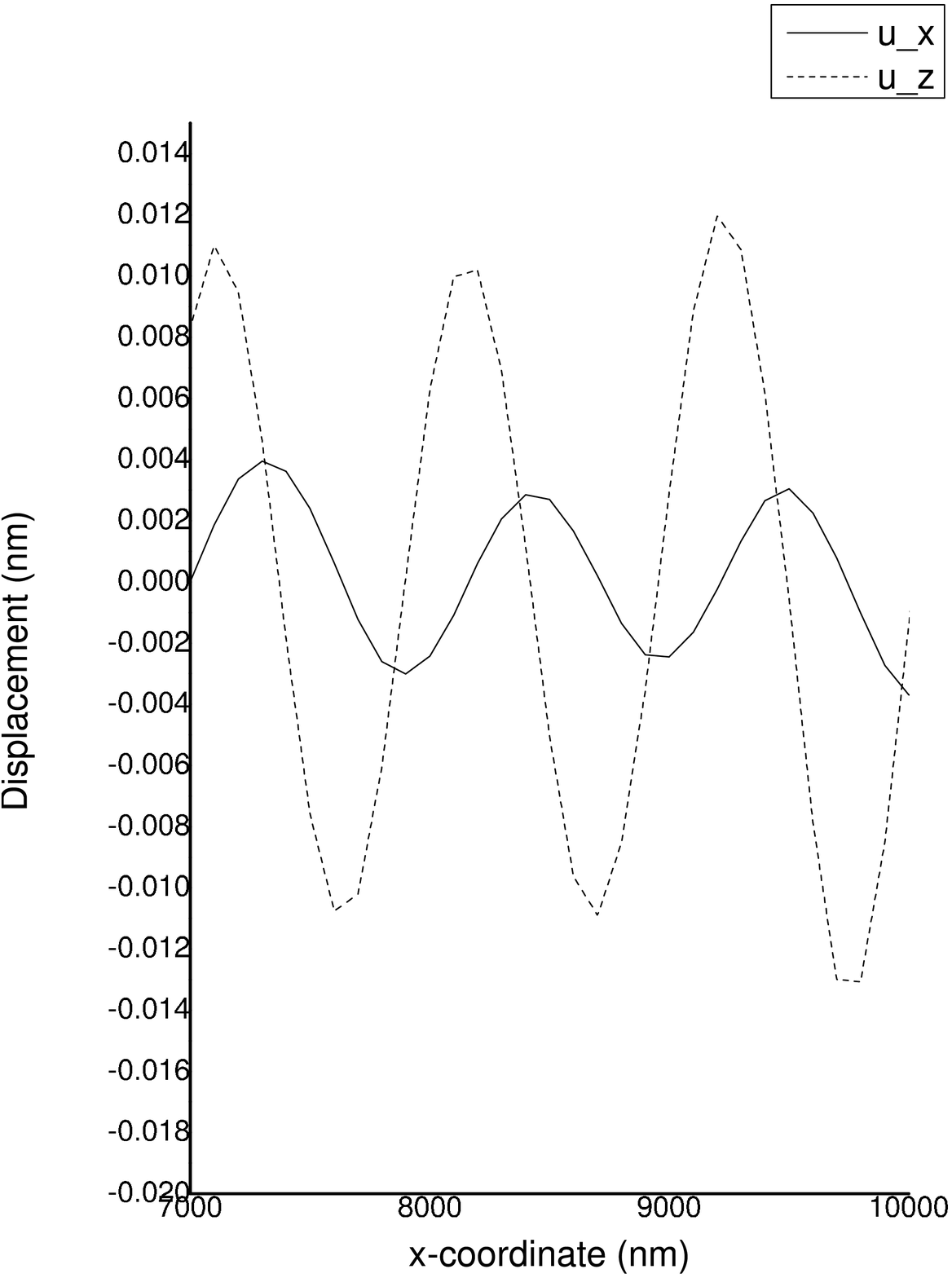}}
\end{center}
\caption{The relative amplitudes and phases of the
displacements $u_x$ and $u_z$, parallel and perpendicular respectively,
to the SAW
propagation.} \label{fig:phases}
\end{figure}

\begin{figure}
\begin{center}

\rotatebox{0}{\scalebox{0.8}{\includegraphics*[1.5in,0.05in][6.5in,6.5in]{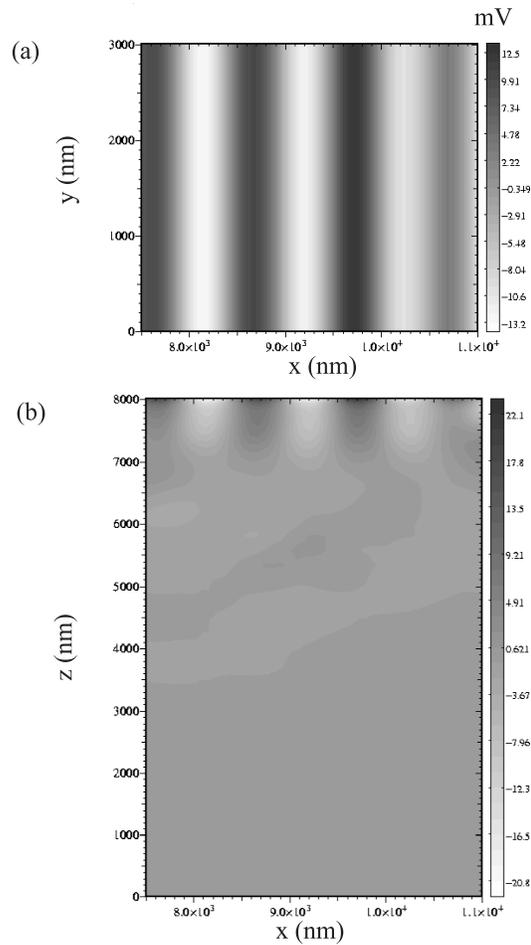}}}
\caption{(a) Plane waves of the electric potential, $100$~nm below
the surface, at time 1.6~ns. (b) Surface nature of the electric
potential, at time 1.6~ns. The SAWs have a significant amplitude
between $z = 8000$~nm and $z = 7000$~nm and have a significantly
reduced amplitude below $z = 7000$~nm} \label{fig:phi_slicesa}
\end{center}
\end{figure}

\begin{figure}[h]
\begin{center}
\rotatebox{0}{\scalebox{0.4}{\includegraphics*[0.25in,0.25in][6.0in,4.5in]{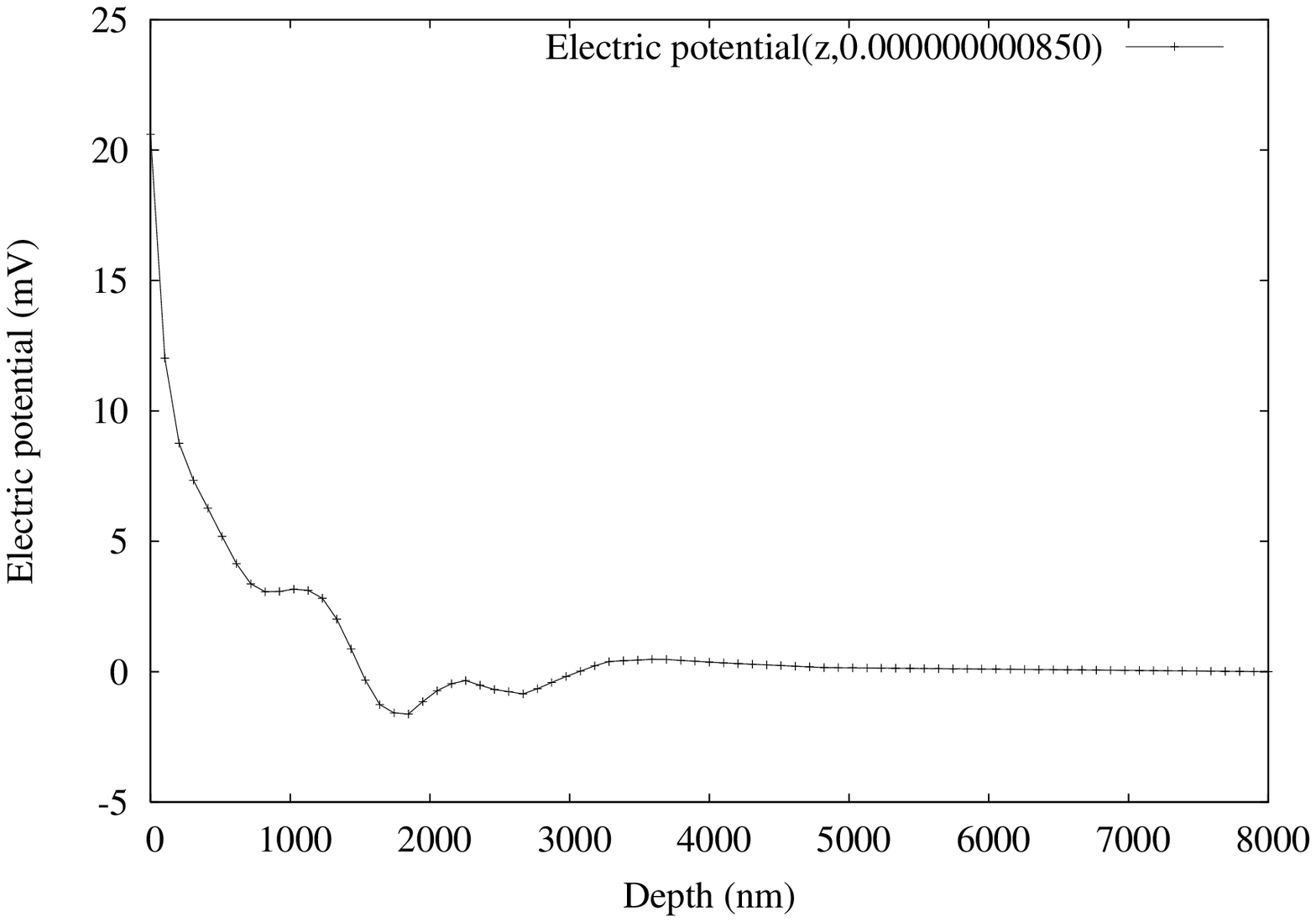}}}
\end{center}
\caption{A typical curve of the electric potential as a function
of depth into the bulk. The precise shape of the curve clearly
depends on the exact time and spatial coordinate the data was
taken as the amplitude of the SAW varies from $-20$~mV to
$+20$~mV. However, all such curves have the oscillatory decay into
the bulk becoming negligible within a few microns.}
\label{fig:phiz}
\end{figure}

Figure~\ref{fig:phases} is a result of a simulation showing the
$\frac{\pi}{2}$ phase difference between the $x$ and $z$
components of the displacement vector, and the larger $z$
amplitude, suggesting elliptical polarization in the sagittal
plane. Figures~\ref{fig:phi_slicesa}(a) and
~\ref{fig:phi_slicesa}(b) show two-dimensional slices, in
perpendicular planes, of the full three-dimensional solutions. The
first shows the non-decaying nature of these waves along the
propagation direction. The SAW wavelength can be seen to be
approximately 1000~nm and its velocity is computed to be
approximately 2700 $\pm 20$~ms$^{-1}$, which is in good agreement
with analytical calculations for the wave velocity. The second
figure shows the surface nature of these waves; the greatest
amplitude is near the surface and there is no observable decay in
the direction of propagation but they decay exponentially into the
bulk. On the far left of each image (close to where the boundary
condition is applied), bulk waves may be observed but as the waves
propagate towards the right, they dissipate all their energy into
the bulk and the only remaining waves are the surface waves.
Figure~\ref{fig:phiz} shows the SAW amplitude as a function of
depth. We see that the amplitude undergoes oscillatory (complex
exponential) decay into the bulk and is negligible a few microns
below the surface, again in good agreement with analytical
expressions derivable from Eq.~\refeq{eqn:SAW} \cite{Simon}.

\subsubsection{The effect of the compound mechanical structure.}

%\begin{widetext}
\begin{figure}[h]
\rotatebox{0}{\scalebox{0.8}{\includegraphics*[0.25in,0.0in][5.5in,5.75in]{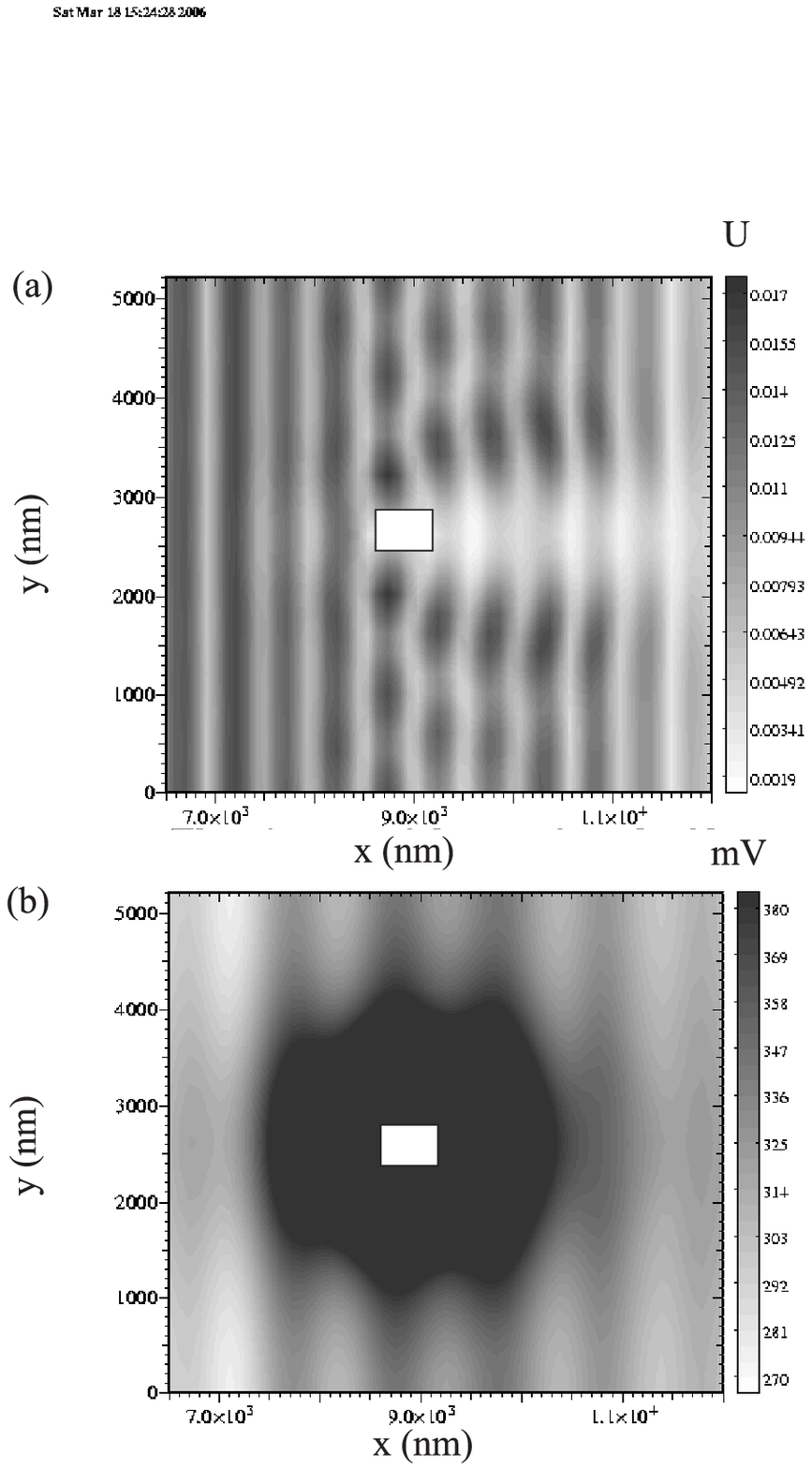}}}

\caption{(a) The magnitude of the displacement field $U=\sqrt{u_r
u_r}$ , at time 2~ns, as the SAW moving from left to right, passes
through the gate centered at ($8.9\cdot 10^{3}$~nm, $2.6\cdot
10^{3}$~nm) and shown by a white square. The mechanical motion is
decoupled from the electric field. (b) The total electric
potential, at time 2~ns, as the SAW passes through the gate when
the mechanical motion is decoupled from the electric field. The
scale has been restricted so that the SAW potential is visible.
The central blur, which is due to the large external potential has
the value of 1.5~V at its center. The white square shows the
position of the gate.} \label{fig:u_mag_C_Dec_xy}
\end{figure}
%\end{widetext}

Figure~\ref{fig:u_mag_C_Dec_xy}(a) shows the magnitude of the
displacement field as the SAW moving from the left of the figure
to the right, passes through the gate when the mechanical motion
is decoupled from the electric field. The gate is centered at
($8.9\cdot 10^{3}$~nm, $2.6\cdot 10^{3}$~nm). We observe a peak to
the left of the gate (due to a reflection) and a trough (due to
damping of the wave) to the right. We also just barely see
vibrations moving away from the gate along the $y$ direction.
Since in this simulation, the mechanical motion is independent of
the electric fields, this damping is due purely to the presence of
a mechanical structure on the surface. The damping of the
mechanical amplitude may be large enough to affect the electric
potential. In this simulation, the electric potential is not
affected significantly as shown in
Fig.~\ref{fig:u_mag_C_Dec_xy}(b). The central blur is the electric
potential due to the gate. The SAW peak emerging from the gate has
been damped in the central region.

\subsubsection{The effect of electromechanical coupling.}

%\begin{widetext}
\begin{figure}[!h]
\rotatebox{0}{\scalebox{0.8}{\includegraphics*[0.25in,0.1in][5.5in,6.0in]{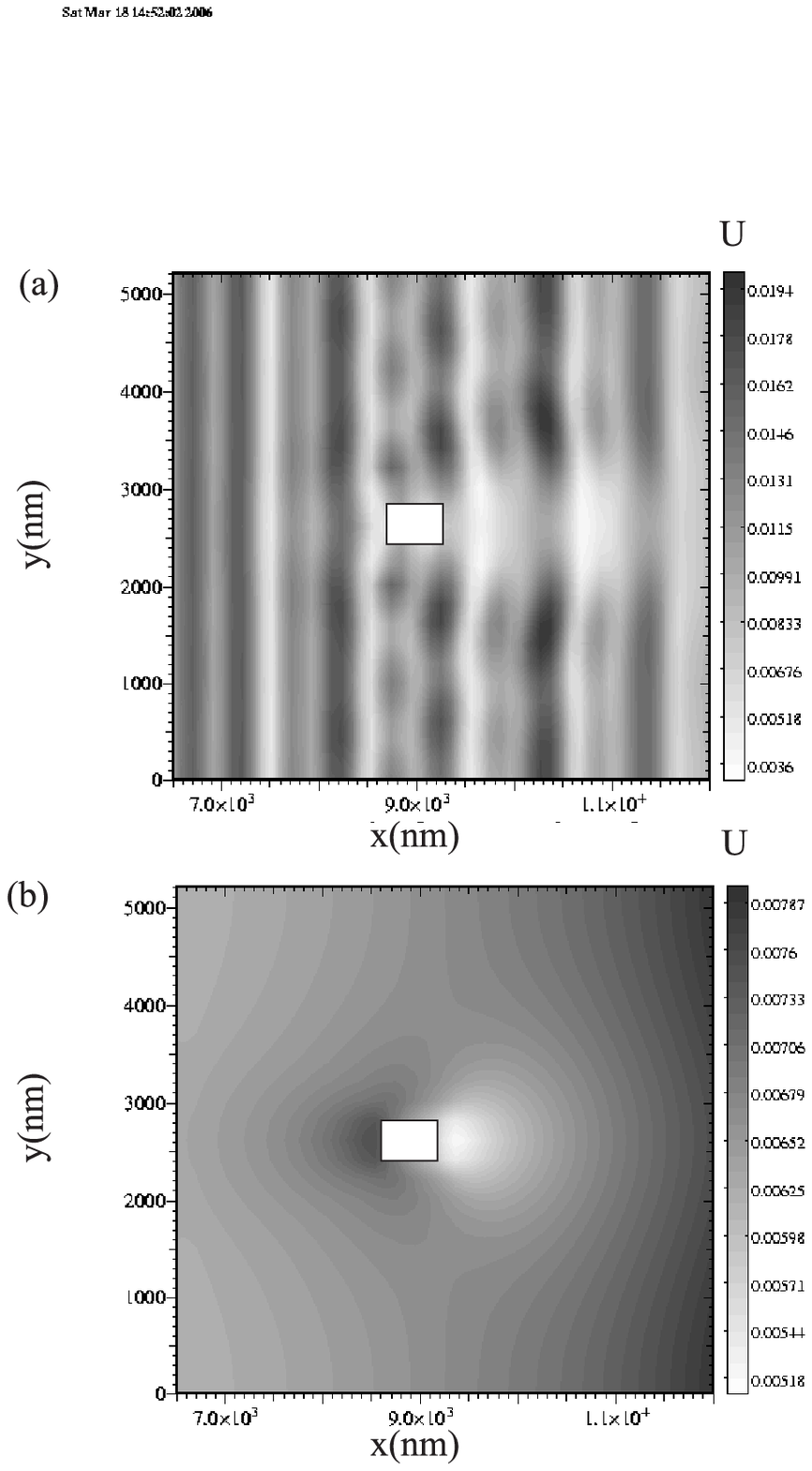}}}

\caption{(a) The magnitude of the displacement field $U=\sqrt{u_r
u_r}$ , at time 2~ns, as the SAW moving from left to right, passes
through the gate  centered at ($8.9\cdot 10^{3}$~nm, $2.6\cdot
10^{3}$~nm) and shown by a white square, when full coupling
between electric and mechanical fields is allowed. (b) The
magnitude of the displacement field $U=\sqrt{u_r u_r}$, at time 0,
demonstrating the mechanical strains caused purely by the gate
with a 1.5~V applied voltage centered at ($8.9\cdot 10^{3}$~nm,
$2.6\cdot 10^{3}$~nm) and shown by a white square, when full
coupling between electric and mechanical fields are allowed.}
\label{fig:u_mag_C_Cpl_xy}
\end{figure}
%\end{widetext}

Figure~\ref{fig:u_mag_C_Cpl_xy}(a) shows the magnitude of the
displacement field, at time 2~ns, as the SAW passes through the
gate, in the case where mutual coupling between the electric and
mechanical fields is allowed. As in the decoupled case, we observe
a peak to the left of the gate (due to reflection of the wave) and
a trough (due to damping of the wave) to the right. Again, we also
see vibrations moving away from the gate along the $y$ direction.
Comparing Figs.~\ref{fig:u_mag_C_Dec_xy}(a) and
\ref{fig:u_mag_C_Cpl_xy}(a), we see that there is additional
mechanical deformation throughout the material due to the presence
of a charged metallic gate. Also, the scale of
Fig.~\ref{fig:u_mag_C_Cpl_xy}(a) is shifted up  compared to that
of Fig.~\ref{fig:u_mag_C_Dec_xy}(a). By looking at the
displacement field at time 0 i.e. before the arrival of the SAW,
we can see the displacement resulting from the applied electric
field. This is shown in Fig.~\ref{fig:u_mag_C_Cpl_xy}(b).

\clearpage
\section{Summary and concluding remarks} \label{sec:Summary}

Our aim with this paper is to suggest a computationally fast
method for simulating SAWs in piezoelectric devices where stress,
deformation, and a quasi-static electric field are fully coupled.
The basic idea of the numerical scheme is to use a finite
difference approximation in time that decouples the elasticity and
the electric field problems such that these can be solved in
sequence at each time level. This decoupling can also be explored
in computer implementations, because independent solvers for
anisotropic elasticity problems and Poisson problems can be joined
together. We showed in particular how object-oriented programming
techniques can realize such couplings in a very convenient way.

Through numerical experiments in a physically one-dimensional wave
propagation problem we have verified that the three-dimensional
code reproduces the expected quadratic convergence in space and
time if linear or trilinear elements are used. Finally, we have
applied the proposed numerical methodology to real SAW phenomena
in a device with complicated geometry. We have described a system
of boundary conditions which are capable of exciting SAW modes in
a small computational domain. We have performed simulations where
we have demonstrated the effects on the SAW of a charged metallic
gate. The results indicate that the method is capable of
predicting the expected complex elasto-electric dynamics in such a
device.

The methodology could easily be applied to simulate SAWs through
devices such as GaAs/AlGaAs heterostructures with more complicated
surface gate geometries.

The principal limitation of the equation splitting is a stability
criterion on the time step length. Implicit methods may remove
time step restrictions, but at a cost of the need to solve large
coupled linear systems at each time level. The computer
implementation is also more involved and requires a
special-purpose solution rather than just joining two well-tested
equation components.  However, for wave propagation problems one
usually needs a fine mesh and a small time step to resolve the
waves, typically leading to $h/\Delta t =\hbox{const}$, which is
in accordance with the stability criterion. Explicit time stepping
approaches are therefore highly relevant and allow efficient
algorithms and implementations.

Forthcoming numerical work will focus on domain decomposition
methods for parallelizing the solver and thereby enable simulation
of large-scale SAW problems. The described time stepping approach
and associated equation splitting are particularly well suited for
parallel computing, because the elasticity problem can be made
``perfectly parallel'', and very efficient parallelization
strategies exist for the Poisson equation.

\section{Acknowledgements}

We thank X. Cai and A. M. Bruaset of Simula Research Laboratory
for providing assistance in the programming and debugging of the
code, and M. Kataoka of the Cavendish Laboratory for useful
discussions. We also acknowledge the Cambridge-MIT Institute,
Darwin College Cambridge, and the Simula Research Laboratory for
financial support.

%\bibliographystyle{unsrt}
%\bibliographystyle{plain}
%\bibliography{bib_SR}

\end{document}